\newcounter{myctr}
\def\myitem{\refstepcounter{myctr}\bibfont\noindent\ifnum\themyctr>9\else\phantom{0}\fi\hangindent17pt\themyctr.\enskip}

\documentclass{ws-ijqi}

\usepackage{braket}

\begin{document}

\markboth{Takuya Machida}
{A quantum walk with a delocalized initial state: contribution from a coin-flip operator}

%%%%%%%%%%%%%%%%%%%%% Publisher's Area please ignore %%%%%%%%%%%%%%
%\catchline{}{}{}{}{}
%%%%%%%%%%%%%%%%%%%%%%%%%%%%%%%%%%%%%%%%%%%%%%%%%%%%%%%%%%%%%%%%%%%

\title{A quantum walk with a delocalized initial state:\\contribution from a coin-flip operator}

\author{Takuya Machida}

\address{Japan Society for the Promotion of Science, Japan\\
Department of Mathematics, University of California, Berkeley, USA\\
machida@stat.t.u-tokyo.ac.jp}

\maketitle

\begin{history}
%\received{Day Month Year}
%\revised{Day Month Year}
%\accepted{Day Month Year}
%\comby{(xxxxxxxxxx)}
\end{history}

\begin{abstract}
A unit evolution step of discrete-time quantum walks is determined by both a coin-flip operator and a position-shift operator.
The behavior of quantum walkers after many steps delicately depends on the coin-flip operator and an initial condition of the walk.
To get the behavior, a lot of long-time limit distributions for the quantum walks starting with a localized initial state have been derived.
In the present paper, we compute limit distributions of a 2-state quantum walk with a delocalized initial state, not a localized initial state, and discuss how the walker depends on the coin-flip operator.
The initial state induced from the Fourier series expansion, which is called the $(\alpha,\beta)$ delocalized initial state in this paper, provides different limit density functions from the ones of the quantum walk with a localized initial state.
\end{abstract}


%%%%%%%%%%%%%%%%%%%%%%%%%%%%%%%%%%%   INTRODUCTION   %%%%%%%%%%%%%%%%%%%%%%%%%%%%%%%%%%%%%%%%%%%%%%%%%%%%%%%%%
\section{Introduction}

Discrete-time 2-state quantum walks (QWs) are considered as motions of quantum particles with the up-spin or down-spin state.
The system of quantum walkers is described by probability amplitude, and its dynamics is determined by a unitary operator which is constructed of both a coin-flip operator and a position-shift operator.
That is, the QW is a unitary process.
One of the goals of the study is to know where the quantum walker is after many steps.
To find the answer, a lot of long-time limit theorems have been derived.
Particularly, the limit distributions play an important role in letting us know the spatial distribution of the walkers after long time.
While we theoretically investigate the QWs, experiments on the QW after some steps have been also realized in physical quantum systems recently.
In anticipation of successful realizations of QWs with many steps, we require investigation of limit theorems which can justify the results of the experiments.

The QWs have provided us with impressive limit distributions. 
The limit distribution of the 2-state QWs on the line $\mathbb{Z}=\left\{0,\pm 1,\pm2,\ldots\right\}$ is depicted by a density function with a compact support like the arcsine law~\cite{Konno2002,Konno2005}. 
If the quantum walkers possess more than three states, they can spatially localize in probability distribution and the limit density functions are described by both a continuous function and the Dirac $\delta$-function which means occasion of localization~\cite{InuiKonnoSegawa2005,InuiKonno2005,SegawaKonno2008,KonnoMachida2010}.
The limit density functions of the QW on the two-dimensional square lattice also have a compact support~\cite{WatabeKobayashiKatoriKonno2008,DiMcMachidaBusch2011}.
Questions concerning the long-time asymptotic behavior of the QW on a three-dimensional lattice currently remain open.
Besides lattices, the Grover walks, whose coin-shift operators are defined by symmetric unitary matrices, were analyzed on trees, joint half lines or a spidernet~\cite{ChisakiHamadaKonnoSegawa2009,ChisakiKonnoSegawa2010,KonnoObataSegawa2013}.
Recently, Liu and Petulante~\cite{LiuPetulante2012} got a weak limit theorem for the QW on the half-line.
While these limit theorems are results for the QWs with a localized initial condition, we will treat QWs starting with a delocalized initial condition in this paper.
To understand properties of the QWs, we need to clarify the relation between the QWs and more general initial conditions.
In fact, when we study quantum systems, we sometimes focus on initial conditions distributed over a domain.
For example, in quantum search algorithms (e.g. Grover's search algorithm), we use quantum systems starting from a delocalized initial state. 
So, the aim of this paper is to analyze the QW on the line with a delocalized initial state, and we will report contribution from the coin-flip operator to the limit distributions.
When the walker on the line starts with a localized initial state, it is known that the coin-flip operator gives an effect to asymmetric shape of the limit distribution~\cite{Konno2002,Konno2005}.
It would be interesting to explore the dependence on the coin-flip operator in the case of a delocalized initial state.

Although the initial state given in this paper is special, our results would become the first stage to capture the relation between the limit distribution of the QW with a delocalized initial state and the coin-flip operator, not numerically but rigorously. 
By knowing how the QW depends on the initial state, we can reach deep understanding of its interesting features.
As a result, the QWs might become attractive in a field of application and spread into science more widely.
There are a few results for the QWs distributed widely in an initial state~\cite{AbalDonangeloRomanelliSiri2006,AbalSiriRomanelliDonangelo2006,ValcarcelRoldanRomanelli2010,ChandrashekarBusch2012}.
Abal et. al.~\cite{AbalDonangeloRomanelliSiri2006,AbalSiriRomanelliDonangelo2006} analyzed a QW starting from two positions.
Chandrashekar and Busch~\cite{ChandrashekarBusch2012} reported numerical results for a QW with an initial state ranging over an area.
Valc{\'a}rcel et al.~\cite{ValcarcelRoldanRomanelli2010} treated the QW initialized by a Gaussian-like distribution in a continuum limit.

The properties of QWs have been clarified theoretically, while there are some experimental results of QWs, which were performed by trapped ions, atoms in optical lattices, photons, optical devices and Bose-Einstein condensates
~\cite{BouwmeesterMarzoliKarmanSchleichWoerdman1999,BroomeFedrizziLanyonKassalAspuru-GuzikWhite2010,DoStohlerBalasubramanianElliottEashFischbachFischbachMillsZwickl2005,KarskiForsterChoiSteffenAltMeschedeWidera2009,PeretsLahiniPozziSorelMorandottiSilberberg2008,SchmitzMatjeschkSchneiderGlueckertEnderleinHuberSchaetz2009,SchreiberCassemiroPotovcekGabrisMosleyAnderssonJexSilberhorn2010,ZahringerKirchmairGerritsmaSolanoBlattRoos2010,MeineckePouliosPolitiMatthewsPeruzzoIsmailWorhoffOBrienThompson2013,SansoniSciarrinoValloneMataloniCrespiRamponiOsellame2012}.
Although a sizable number of steps in random walks are easily realized, much effort is required to instantiate many steps in QWs.
For example, Z\"{a}hringer et al.~\cite{ZahringerKirchmairGerritsmaSolanoBlattRoos2010} implemented a QW with up to 23 steps by using trapped ions.
Since realization of QWs is developing recently, we need to prepare rigorous results to contribute experiments on QWs.

The rest of this paper is organized as follows.
In Sec.~\ref{sec:definition}, we introduce the definition of a discrete-time 2-state QW on the line and the $(\alpha,\beta)$ delocalized initial state which is a wider initial state.
Section~\ref{sec:limit_th} is devoted to showing our limit theorem.
In Sec.~\ref{sec:example}, we show contribution from the coin-flip operator to the limit theorem of the QW with the $(\alpha,\beta)$ delocalized initial state in some examples.
In the final section, we conclude our results and discuss a future problem.

%%%%%%%%%%%%%%%%%%%%%%%%%%%%%%%%%%%   DEFINITION   %%%%%%%%%%%%%%%%%%%%%%%%%%%%%%%%%%%%%%%%%%%%%%%%%%%%%%%%%
\section{A discrete-time QW on the line with a delocalized initial state}
\label{sec:definition}
Total system of discrete-time 2-state QWs on the line is defined in a tensor space $\mathcal{H}_p\otimes\mathcal{H}_c$, where $\mathcal{H}_p$ is called a position Hilbert space which is spanned by a basis $\left\{\ket{x}:\,x\in\mathbb{Z}\right\}$ and $\mathcal{H}_c$ is called a coin Hilbert space which is spanned by a basis $\left\{\ket{0},\ket{1}\right\}$ with the vectors $\bra{0}=[1,0],\,\bra{1}=[0,1]$.
Let $\ket{\psi_{t}(x)} \in \mathcal{H}_c$ be the probability amplitudes of the walker at position $x$ at time $t \in\left\{0,1,2,\ldots\right\}$.
Physical meaning of the component $\braket{0|\psi_t(x)}$ (resp. $\braket{1|\psi_t(x)}$) is the amplitude of the up-spin (resp. down-spin) state of the quantum particle at position $x$ at time $t$. 
The state of the 2-state QWs on the line at time $t$ is expressed by $\ket{\Psi_t}=\sum_{x\in\mathbb{Z}}\ket{x}\otimes\ket{\psi_{t}(x)}$.
Time evolution of our QWs is described by a unitary matrix, that is a coin-flip operator
\begin{align}
 U_{\xi,\theta}=&\cos\theta\ket{0}\bra{0}+(-1)^{\xi}\sin\theta\ket{0}\bra{1}+\sin\theta\ket{1}\bra{0}-(-1)^\xi\cos\theta\ket{1}\bra{1}
\end{align}
with $\xi\in\left\{0,1\right\}$ and $\theta\in [0,2\pi)$.
In this paper, we focus on the case of $\theta\neq \pi/2, \pi, 3\pi/2$.
We should note that $U_{1,\frac{\pi}{2}-\theta}=(\ket{0}\bra{1}+\ket{1}\bra{0})U_{0,\theta}$ and $\det U_{\xi,\theta}=(-1)^{\xi}$.
The amplitudes evolve along
\begin{align}
 \ket{\psi_{t+1}(x)}=&\ket{0}\bra{0}U_{\xi,\theta}\ket{\psi_t(x+1)}+\ket{1}\bra{1}U_{\xi,\theta}\ket{\psi_t(x-1)}.\label{eq:te}
\end{align}
Equation~(\ref{eq:te}) means that the walker's position is shifted after its coin states are flipped.
The probability that the quantum walker $X_t$ can be found at position $x$ at time $t$ is defined by
\begin{align}
 \mathbb{P}(X_t=x)=&\braket{\psi_t(x)|\psi_t(x)}=|\braket{0|\psi_t(x)}|^2+|\braket{1|\psi_t(x)}|^2.
\end{align}
The term $|\braket{0|\psi_t(x)}|^2$ (resp. $|\braket{1|\psi_t(x)}|^2$) denotes the probability that the quantum particle with the up-spin (resp. down-spin) state can be found at position $x$ at time $t$.
Given the initial states $\ket{\psi_0(x)}$, we get the probability distribution $\mathbb{P}(X_t=x)$ for any time $t$.
In this paper, we focus on a special delocalized initial state as follows.

%{\bf Our delocalized initial state is constructed by using the Fourier series expansion.}
Let $w_1,w_2:\mathbb{R}\longrightarrow\mathbb{R}$ be the functions that satisfy $w_j(k+2\pi)=w_j(k)\,(j=1,2)$ and $w_j\in L^2[-\pi,\pi]$.
Assuming $\int_{-\pi}^{\pi}\left\{|w_1(k)|^2+|w_2(k)|^2\right\}dk > 0$, we set the following initial condition on the QW,
\begin{equation}
 \ket{\psi_0(x)}=\frac{1}{\sqrt{W(w_1,w_2)}}\left\{d_1(x)+id_2(x)\right\}\ket{\phi},\label{eq:upis}
\end{equation}
where
\begin{align}
 W(w_1,w_2)=&\int_{-\pi}^{\pi}\left\{|w_1(k)|^2+|w_2(k)|^2\right\}dk,\\
 d_j(x)=&\frac{1}{\sqrt{2\pi}}\int_{-\pi}^{\pi} w_j(k)e^{ikx}dk\quad(j=1,2),
\end{align}
and $\ket{\phi}=\alpha\ket{0}+\beta\ket{1}$ with $\alpha,\beta\in\mathbb{C}$ and $|\alpha|^2+|\beta|^2=1$.
We should note that $\sum_{x\in\mathbb{Z}}\mathbb{P}(X_0=x)=\sum_{x\in\mathbb{Z}}\braket{\psi_0(x)|\psi_0(x)}=1$ by Parseval's theorem.
Since Eq.~(\ref{eq:upis}) provides a delocalized initial state, we call it the {\it $(\alpha,\beta)$ delocalized initial state} in this paper.

%%%%%%%%%%%%%%%%%%%%%%%%%%%%%%%%%%%   RESULT   %%%%%%%%%%%%%%%%%%%%%%%%%%%%%%%%%%%%%%%%%%%%%%%%%%%%%%%%%
\section{Limit distribution}
\label{sec:limit_th}
Limit distributions depict approximate motion of the QW after many steps, just like the relation between random walks and central limit theorems.
To get the behavior of the QW with the $(\alpha,\beta)$ delocalized initial state as $t\to\infty$, we compute a limit distribution by using the Fourier analysis which is one of the standard methods to derive limit theorems of QWs~\cite{GrimmettJansonScudo2004,Machida2011,InuiKonnoSegawa2005,InuiKonno2005,KonnoMachida2010,SegawaKonno2008,ChisakiKonnoSegawaShikano2011,WatabeKobayashiKatoriKonno2008,DiMcMachidaBusch2011}.
Formal expression of the limit distribution was obtained in Grimmett et al.~\cite{GrimmettJansonScudo2004}.
From the description, however, we don't strictly see how both coin-flip operators and delocalized initial states contribute to limit distributions.
To catch the contribution, we progress some calculation in this section.
From Eq.~(\ref{eq:te}), the time evolution of the Fourier transform $\ket{\hat{\Psi}_{t}(k)}=\sum_{x\in\mathbb{Z}} e^{-ikx}\ket{\psi_t(x)}\,(k\in\left[-\pi,\pi\right))$ becomes $\ket{\hat{\Psi}_{t+1}(k)}= \hat U_{\xi,\theta}(k)\ket{\hat{\Psi}_{t}(k)}=\hat U_{\xi,\theta}(k)^t \ket{\hat\Psi_{0}(k)}$ , where $\hat U_{\xi,\theta}(k)=(e^{ik}\ket{0}\bra{0}+e^{-ik}\ket{1}\bra{1})U_{\xi,\theta}$.
The Carleson-Hunt theorem changes the initial state of the Fourier transform,
\begin{align}
 \ket{\hat\Psi_0(k)}=&\sum_{x\in\mathbb{Z}}\Biggl\{\left(\frac{1}{2\pi}\int_{-\pi}^{\pi}w_1(\tilde{k})e^{i\tilde{k}x}d\tilde{k}\right)e^{-ikx}\nonumber\\
 &+i\left(\frac{1}{2\pi}\int_{-\pi}^{\pi}w_2(\tilde{k})e^{i\tilde{k}x}d\tilde{k}\right)e^{-ikx}\Biggr\}\sqrt{\frac{2\pi}{W(w_1,w_2)}}\ket{\phi}\nonumber\\
 =&\sqrt{\frac{2\pi}{W(w_1,w_2)}}\left\{w_1(k)+iw_2(k)\right\}\ket{\phi},\label{eq:fourierexpand}
\end{align}
almost everywhere on $[-\pi,\pi]$~\cite{JorsboeMejlbro1982,Reyna2002}.
Equation~(\ref{eq:fourierexpand}) is essentially the Fourier series expansions of the functions $w_1(k),w_2(k)$.
After using the method treated in Grimmett et al.~\cite{GrimmettJansonScudo2004}, for $r=0,1,2,\ldots$, we get
\begin{align}
 &\lim_{t\to\infty}\mathbb{E}\biggl[\biggl(\frac{X_t}{t}\biggr)^r\biggr]\nonumber\\
 =&\int_{0}^{\pi} \frac{dk}{2\pi} h(k)^r\,\biggl[\biggl\{\left|\bra{v(k)}J_\xi\ket{\hat\Psi_0(k+\xi\pi/2)}\right|^2+\left|\bra{v(-k)}J_\xi\ket{\hat\Psi_0(-k+\xi\pi/2)}\right|^2\biggr\}\nonumber\\
 &+(-1)^r\biggl\{\left|\overline{\bra{v(\pi-k)}}J_\xi \ket{\hat\Psi_0(k+\xi\pi/2)}\right|^2+\left|\overline{\bra{v(\pi+k)}}J_\xi \ket{\hat\Psi_0(-k+\xi\pi/2)}\right|^2\biggr\}\biggr],
\end{align}
where
\begin{align}
 h(k)=&\frac{c\cos k}{\sqrt{1-c^2\sin^2 k}},\\
 \ket{v(k)}=&\frac{e^{ik}s}{\sqrt{N(k)}}\ket{0}-\frac{c\cos k+\sqrt{1-c^2\sin^2 k}}{\sqrt{N(k)}}\ket{1},\\
 N(k)=&1+s^2+c^2\cos 2k+2c\cos k\sqrt{1-c^2\sin^2 k},\\
 J_\xi=&\ket{0}\bra{0}+(-1)^\xi\ket{1}\bra{1},
\end{align}
and $c=\cos\theta, s=\sin\theta$.
By putting $h(k)=x$, we have
\begin{align}
 \lim_{t\to\infty}\mathbb{E}\left[\left(\frac{X_t}{t}\right)^r\right]=
  \int_{-\infty}^\infty x^r &\frac{|s|}{2\pi(1-x^2)\sqrt{c^2-x^2}}\eta(x)I_{(-|c|,|c|)}(x)\,dx,\label{eq:limit}
\end{align}
where $I_A(x)=1$ if $x\in A$, $I_A(x)=0$ if $x\notin A$ and
 \begin{align}
  \eta(x)=&\left|\bra{v(\kappa(x))}J_\xi\ket{\hat\Psi_0(\kappa(x)+\xi\pi/2)}\right|^2\nonumber\\
  &+\left|\bra{v(\kappa(x))}J_\xi\ket{\hat\Psi_0(\kappa(x)-\pi+\xi\pi/2)}\right|^2\nonumber\\
  &+\left|\bra{v(-\kappa(x))}J_\xi\ket{\hat\Psi_0(-\kappa(x)+\xi\pi/2)}\right|^2\nonumber\\
  &+\left|\bra{v(-\kappa(x))}J_\xi\ket{\hat\Psi_0(\pi-\kappa(x)+\xi\pi/2)}\right|^2,\\
  \kappa(x)=&\arccos\left(\frac{|s|x}{c\sqrt{1-x^2}}\right)\,\in\left[0,\pi\right].
 \end{align}
In Eq.~(\ref{eq:limit}), if we give a special initial state to the Fourier transform $\ket{\hat\Psi_0(k)}$, we can calculate a limit distribution.
Let $F:\mathbb{R}\longrightarrow\mathbb{C}$ be a function that satisfies $F(k+2\pi)=F(k)$, $\int_{-\pi}^{\pi}\left|F(k)\right|^2 dk=2\pi$ and $\Re(F(k)),\,\Im(F(k))\in C^{\infty}[-\pi,\pi]$ almost everywhere, where $\Re(z)$ (resp. $\Im(z)$) denotes the real (resp. imaginary) part of $z\in\mathbb{C}$ and $\mathbb{R}\, (resp.\, \mathbb{C})$ means the set of real (resp. complex) numbers.
If we set $\ket{\hat\Psi_0(k)}=F(k)\ket{\phi}$, we obtain
\begin{align}
 &\lim_{t\to\infty}\mathbb{E}\biggl[\left(\frac{X_t}{t}\right)^r\biggr]\nonumber\\
 =&\int_{-\infty}^\infty x^r\, \Bigl\{f_1(x;\xi,\alpha,\beta)\eta_1(x;\xi)+f_2(x;\xi,\alpha,\beta)\eta_2(x;\xi)\Bigr\}I_{(-|c|,|c|)}(x)\,dx,\label{eq:limit_th}
\end{align}
where
\begin{align}
 f_1(x;\xi,\alpha,\beta)=&\frac{|s|}{\pi(1-x^2)\sqrt{c^2-x^2}}\biggl[1-\biggl\{|\alpha|^2-|\beta|^2+(-1)^\xi\frac{2s\Re(\alpha\overline{\beta})}{c}\biggr\}x\biggr],\\
 f_2(x;\xi,\alpha,\beta)=&-(-1)^\xi\frac{s\Im(\alpha\overline{\beta})}{|c|\pi(1-x^2)},\\
 \eta_1(x;\xi)=&\frac{1}{4}\Bigl\{|F(\kappa(x)+\xi\pi/2)|^2+|F(-\kappa(x)+\xi\pi/2)|^2\nonumber\\
 &+|F(\kappa(x)-\pi+\xi\pi/2)|^2+|F(\pi-\kappa(x)+\xi\pi/2)|^2\Bigr\},\\
 \eta_2(x;\xi)=&\frac{1}{2}\Bigl\{|F(\kappa(x)+\xi\pi/2)|^2-|F(-\kappa(x)+\xi\pi/2)|^2\nonumber\\
 &+|F(\kappa(x)-\pi+\xi\pi/2)|^2-|F(\pi-\kappa(x)+\xi\pi/2)|^2\Bigr\}.
\end{align}
As we can see from Eq.~(\ref{eq:fourierexpand}), the function $F(k)$ corresponds to $\sqrt{\frac{2\pi}{W(w_1,w_2)}}\left\{w_1(k)+iw_2(k)\right\}$.
%, where the functions $w_1(k), w_2(k)$ are given in Sec.~\ref{sec:definition}.
In the next section, we will show contribution from the coin-flip operators $U_{\xi,\theta}$ to the limit theorem Eq.~(\ref{eq:limit_th}) when the functions $w_1(k), w_2(k)$ are supplied.

%%%%%%%%%%%%%%%%%%%%%%%%%%%%%%%%%%EXAMPLES%%%%%%%%%%%%%%%%%%%%%%%%%%%%%%%%%%%%%
\section{Contribution from coin-flip operators}
\label{sec:example}

%When the QWs start from the origin and we take a unitary matrix as the coin-flip operator, the operator affects only their asymmetric quality~\cite{Konno2002,Konno2005}.
In this section, to see contribution from the coin-flip operators $U_{\xi,\theta}$, we concretely give the functions $w_1(k),w_2(k)$ on a range with the length $2\pi$ and compute limit density functions of the QW as $t\to\infty$.
Outside the given domains of $w_1(k),w_2(k)$, the functions shall be $2\pi$-periodically expanded on $\mathbb{R}$.

\subsection{Case 1}
\begin{figure}[h]
 \begin{center}
  \begin{minipage}{50mm}
   \begin{center}
    \includegraphics[scale=0.4]{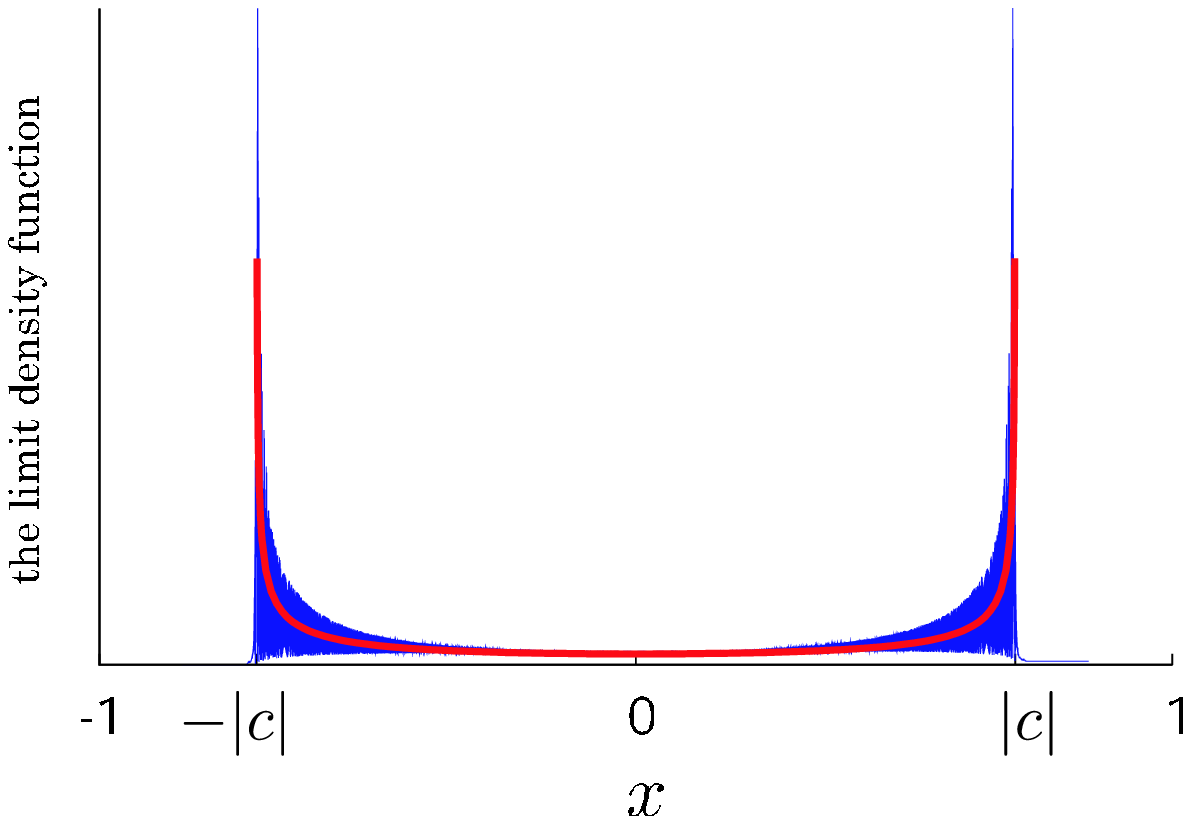}\\
    {(a) $\xi=0$}
   \end{center}
  \end{minipage}
  \begin{minipage}{50mm}
   \begin{center}
    \includegraphics[scale=0.4]{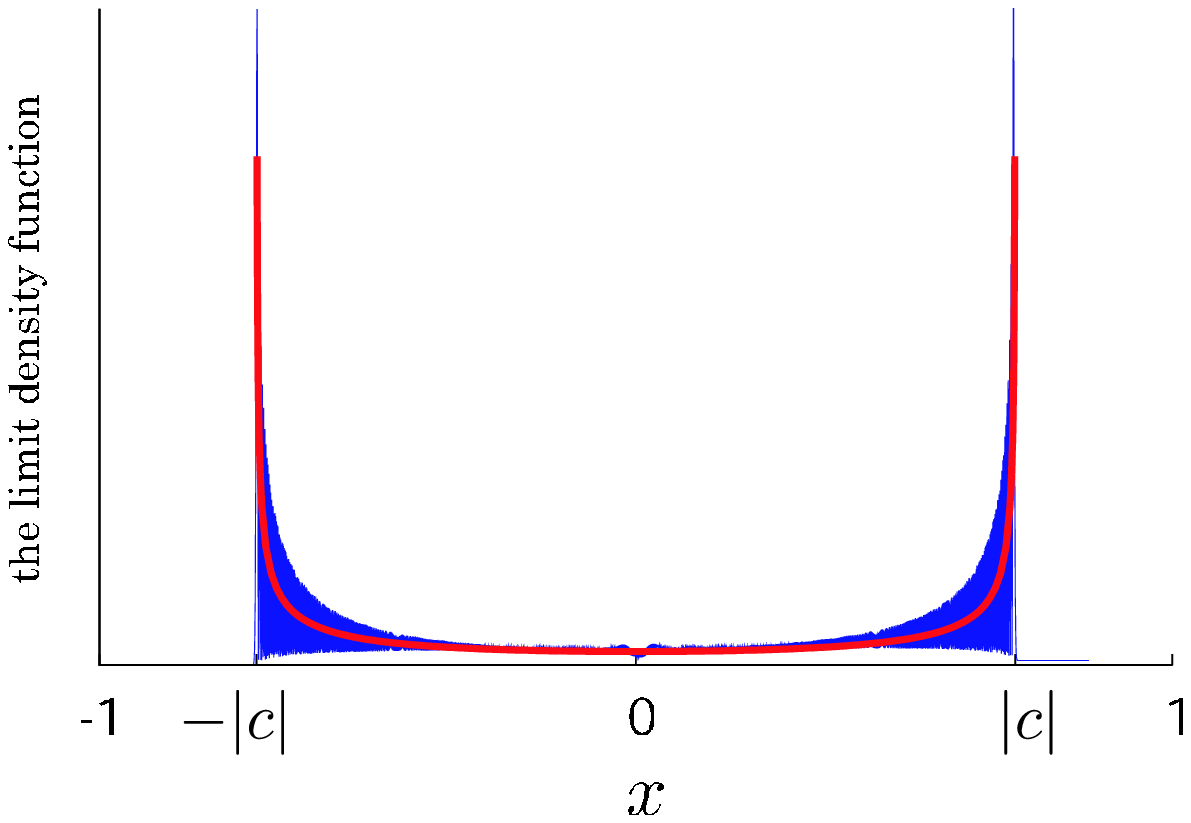}\\
    {(b) $\xi=1$}
   \end{center}
  \end{minipage}
 \end{center}
 \caption{(Case 1) Comparison between the limit density function (red line) and the probability distribution $\mathbb{P}(X_t/t=x)$ at time $t=5000$ (blue line) with $\alpha=1/\sqrt{2},\,\beta=i/\sqrt{2}$.}
 \label{fig:case1}
\end{figure}

We take $w_1(k)=\cos ak\,(-\pi\leq k\leq \pi),\,w_2(k)=\sin ak\,(-\pi<k<\pi)$ with $a\notin \mathbb{Z}$.
Then the $(\alpha,\beta)$ delocalized initial state becomes
\begin{equation}
 \ket{\psi_0(x)}=(-1)^x\frac{\sin a\pi}{(x+a)\pi}\ket{\phi}\quad (x=0,\pm 1,\pm 2,\ldots).\label{eq:initial_case1}
\end{equation}
Before we check the limit distribution for this initial state, we remark the following fact from Eq.~(\ref{eq:limit_th}).
If the function $F(k)$ given in Sec.~\ref{sec:limit_th} satisfies $|F(k-\pi)|=|F(-k)|=|F(k)|$, then we have
\begin{align}
 \lim_{t\to\infty}\mathbb{E}\bigg[\bigg(\frac{X_t}{t}\biggr)^r\biggr]=
  \int_{-\infty}^\infty  x^r f_1(x;\xi,\alpha,\beta)|F(\kappa(x)+\xi\pi/2)|^2 I_{(-|c|,|c|)}(x)\,dx.\label{eq:cor1}
\end{align}
By using this result, we compute the limit distribution of the QW with the initial state Eq.~(\ref{eq:initial_case1}),
\begin{align}
 \lim_{t\to\infty}\mathbb{E}\left[\left(\frac{X_t}{t}\right)^r\right]=&\int_{-\infty}^{\infty}x^r f_1(x;\xi,\alpha,\beta)I_{(-|c|,|c|)}(x)dx\nonumber\\
 =&\int_{-\infty}^{\infty}x^r\,\frac{|s|}{\pi(1-x^2)\sqrt{c^2-x^2}}\,I_{(-|c|,|c|)}(x)\nonumber\\
 &\times\biggl[1-\left\{|\alpha|^2-|\beta|^2+(-1)^\xi\frac{2s\Re(\alpha\overline{\beta})}{c}\right\}x\biggr]\,dx.\label{eq:result_case1}
\end{align}
We can imagine the limit density function with $\alpha=1/\sqrt{2}, \beta=i/\sqrt{2}$ in Fig.~ \ref{fig:case1}.
Equation~(\ref{eq:result_case1}) is the same limit theorem as that of the walker starting from the origin with $\ket{\psi_0(0)}=\ket{\phi}$~\cite{Konno2002,Konno2005}.

\subsection{Case 2}
Given the functions $w_1(k)=\log(4\cos^2 k)\,(\pi/2 < k < 5\pi/2),\,w_2(k)=0$, the initial state is expressed as
\begin{equation}
 \ket{\psi_0(x)}=\left\{\begin{array}{ll}
		  0\ket{\phi}&(x=0,\pm 1,\pm 3,\ldots), \\[1mm]
		  -(-1)^{\frac{x}{2}}\frac{2\sqrt{3}\,|x|}{\pi x^2}\ket{\phi}& (x=\pm 2,\pm 4,\ldots).
			\end{array}\right.
\end{equation}
In a similar fashion as Case 1, we can extract
\begin{align}
 \lim_{t\to\infty}\mathbb{E}\left[\left(\frac{X_t}{t}\right)^r\right]=\int_{-\infty}^{\infty}& x^r f_1(x;\xi,\alpha,\beta)g_1(x;\xi)I_{(-|c|,|c|)}(x)\,dx,\label{eq:limth_case2}
\end{align}
where
\begin{align}
 g_1(x;0)=&\frac{3}{\pi^2}\left\{\log\frac{4s^2x^2}{c^2(1-x^2)}\right\}^2,\\
 g_1(x;1)=&\frac{3}{\pi^2}\left\{\log\frac{4(c^2-x^2)}{c^2(1-x^2)}\right\}^2.
\end{align}

\begin{figure}[h]
 \begin{center}
  \begin{minipage}{50mm}
   \begin{center}
    \includegraphics[scale=0.4]{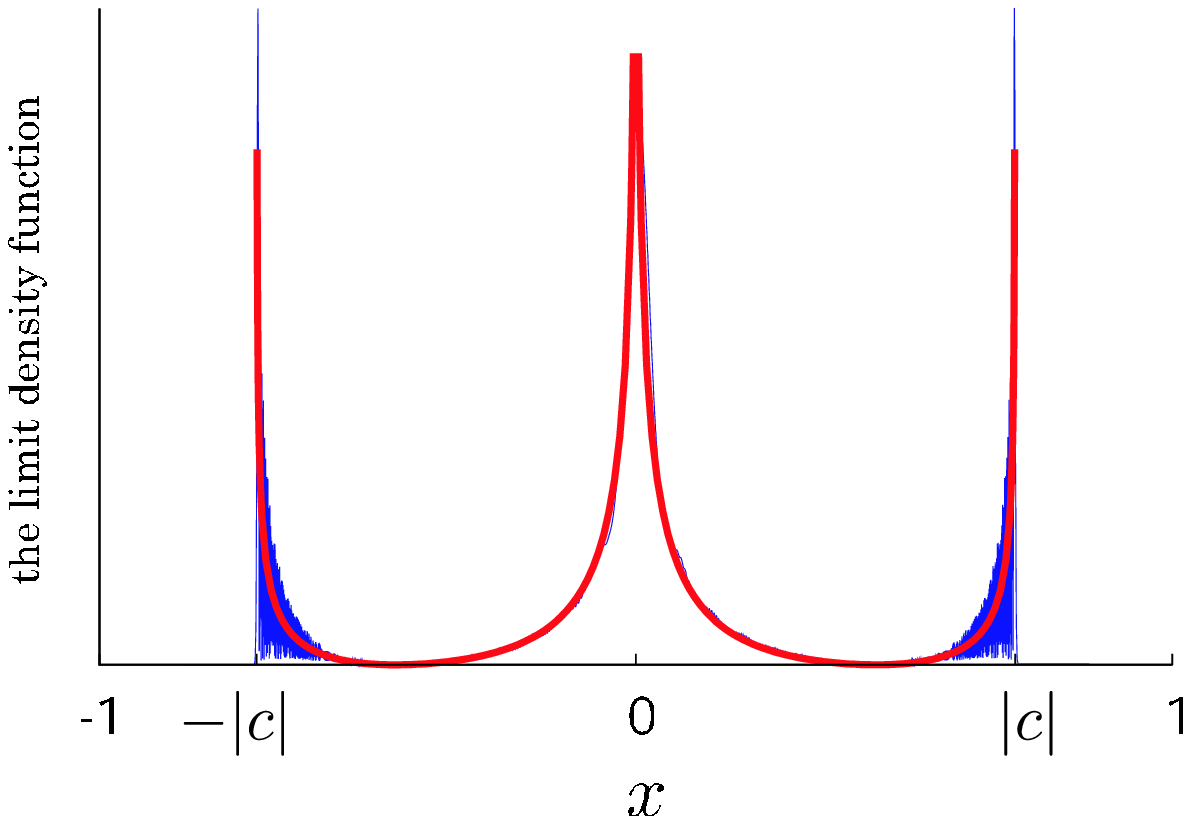}\\
    {(a) $\xi=0$}
   \end{center}
  \end{minipage}
  \begin{minipage}{50mm}
   \begin{center}
    \includegraphics[scale=0.4]{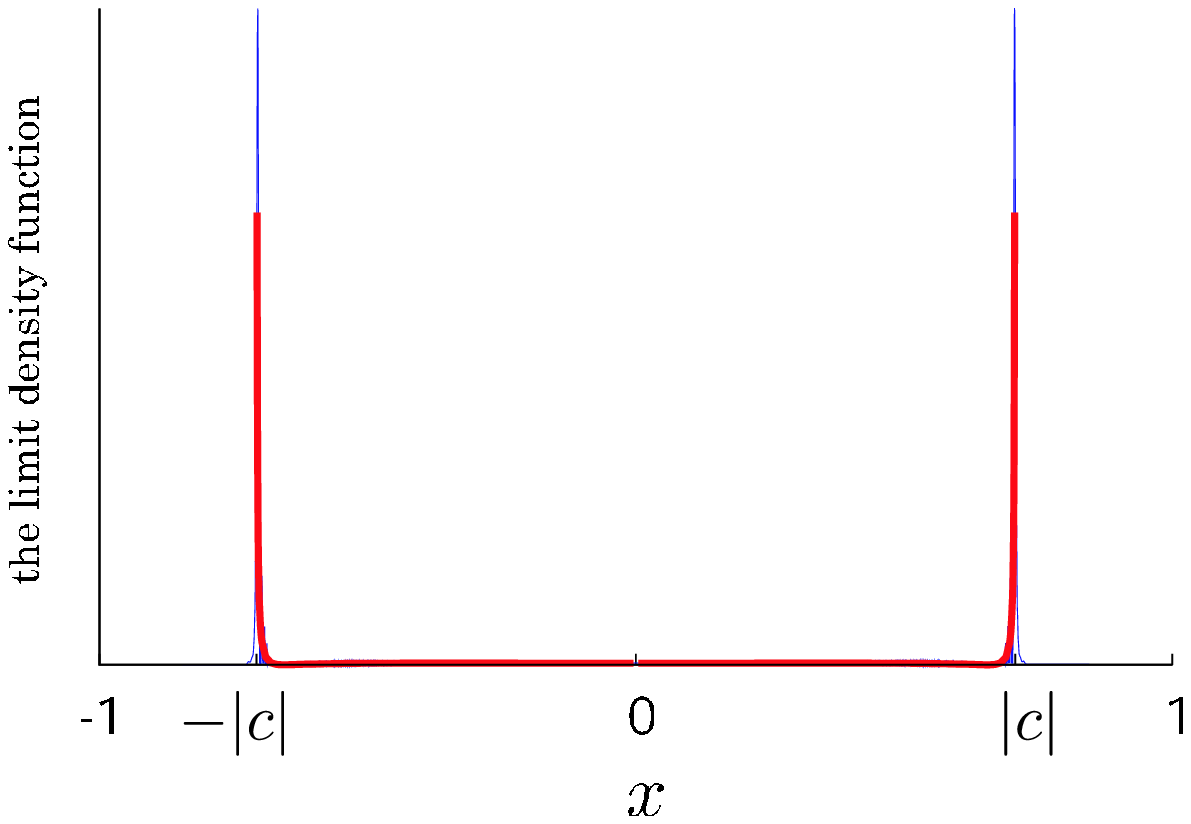}\\
    {(b) $\xi=1$}
   \end{center}
  \end{minipage}
 \end{center}
 \caption{(Case 2) Comparison between the limit density function (red line) and the probability distribution $\mathbb{P}(X_t/t=x)$ at time $t=5000$ (blue line) with $\alpha=1/\sqrt{2},\,\beta=i/\sqrt{2}$.}
 \label{fig:case2}
\end{figure}
\noindent We remark that the limit density function with $\xi=0$ has three singular points $x=0, \pm |c|$.

\subsection{Case 3}

The $(\alpha,\beta)$ delocalized initial state constructed of $w_1(k)=\frac{1}{2}\log(\cot\left|k/2-\pi/4\right|)\,(-\pi/2 < k < 3\pi/2),\,w_2(k)=0$ is written as
\begin{equation}
 \ket{\psi_0(x)}=\left\{\begin{array}{ll}
		  0\ket{\phi}&(x=0,\pm 2,\ldots), \\[1mm]
			 i(-1)^{\frac{|x|-1}{2}}\frac{2}{\pi x}\ket{\phi}& (x=\pm 1,\pm 3,\ldots).
			\end{array}\right.\label{eq:initial_case3}
\end{equation}
The functions $w_1(k),w_2(k)$ don't satisfy the condition corresponding to $|F(k-\pi)|=|F(-k)|=|F(k)|$.
We have to pass through the different computation from Eq.~(\ref{eq:cor1}).
So, we add a result derived immediately from Eq.~(\ref{eq:limit_th}) as follows.
If the function $F(k)$ satisfies $|F(-k)|^2+|F(\pi-k)|^2=|F(k)|^2+|F(k-\pi)|^2$, then we have
\begin{align}
 \lim_{t\to\infty}\mathbb{E}\left[\left(\frac{X_t}{t}\right)^r\right]=
  \int_{-\infty}^\infty  x^r f_1(x;\xi,\alpha,\beta)\eta_3(x;\xi)I_{(-|c|,|c|)}(x)\,dx,\label{eq:cor2}
\end{align}
where
\begin{align}
 \eta_3(x;\xi)=\frac{1}{2}\Bigl\{|F(\kappa(x)+\xi\pi/2)|^2+|F(\kappa(x)-\pi+\xi\pi/2)|^2\Bigr\}.
\end{align}
From Eq.~(\ref{eq:cor2}), we obtain the quantum walker with Eq.~(\ref{eq:initial_case3}) as $t\to\infty$,
\begin{align}
 \lim_{t\to\infty}\mathbb{E}\left[\left(\frac{X_t}{t}\right)^r\right]=\int_{-\infty}^{\infty} & x^r f_1(x;\xi,\alpha,\beta)g_2(x;\xi)I_{(-|c|,|c|)}(x)\,dx.\label{eq:limth_case3}
\end{align}
where
\begin{align}
 g_2(x;0)&=\frac{2}{\pi^2}\Biggl[\left\{\log\left(\frac{|c|\sqrt{1-x^2}+\sqrt{c^2-x^2}}{|sx|}\right)\right\}^2\nonumber\\
 &+\left\{\log\left(\frac{|c|\sqrt{1-x^2}-\sqrt{c^2-x^2}}{|sx|}\right)\right\}^2\Biggr],\\
 g_2(x;1)&=\frac{2}{\pi^2}\Biggl[\left\{\log\left(\frac{c|c|\sqrt{1-x^2}+|cs|x}{c\sqrt{c^2-x^2}}\right)\right\}^2\nonumber\\
 &+\left\{\log\left(\frac{c|c|\sqrt{1-x^2}-|cs|x}{c\sqrt{c^2-x^2}}\right)\right\}^2\Biggr].
\end{align}

\begin{figure}[h]
 \begin{center}
  \begin{minipage}{50mm}
   \begin{center}
    \includegraphics[scale=0.4]{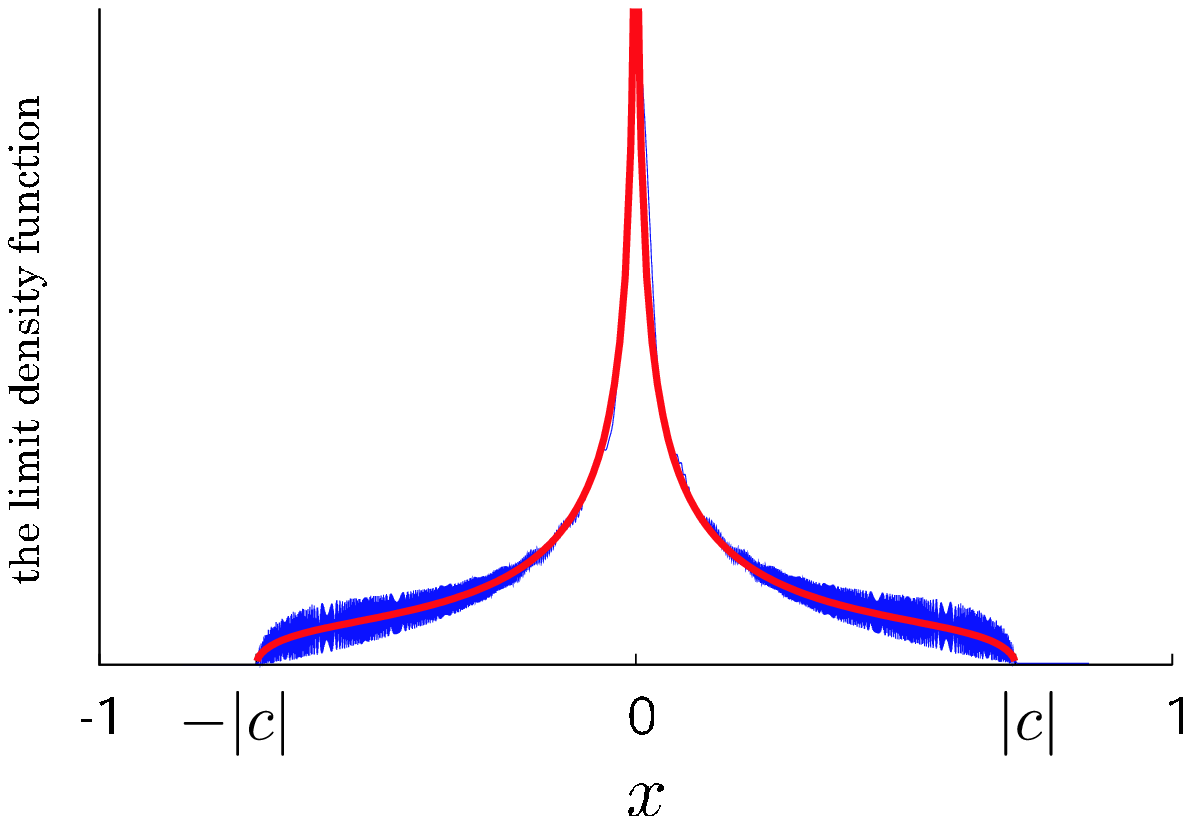}\\
    {(a) $\xi=0$}
   \end{center}
  \end{minipage}
  \begin{minipage}{50mm}
   \begin{center}
    \includegraphics[scale=0.4]{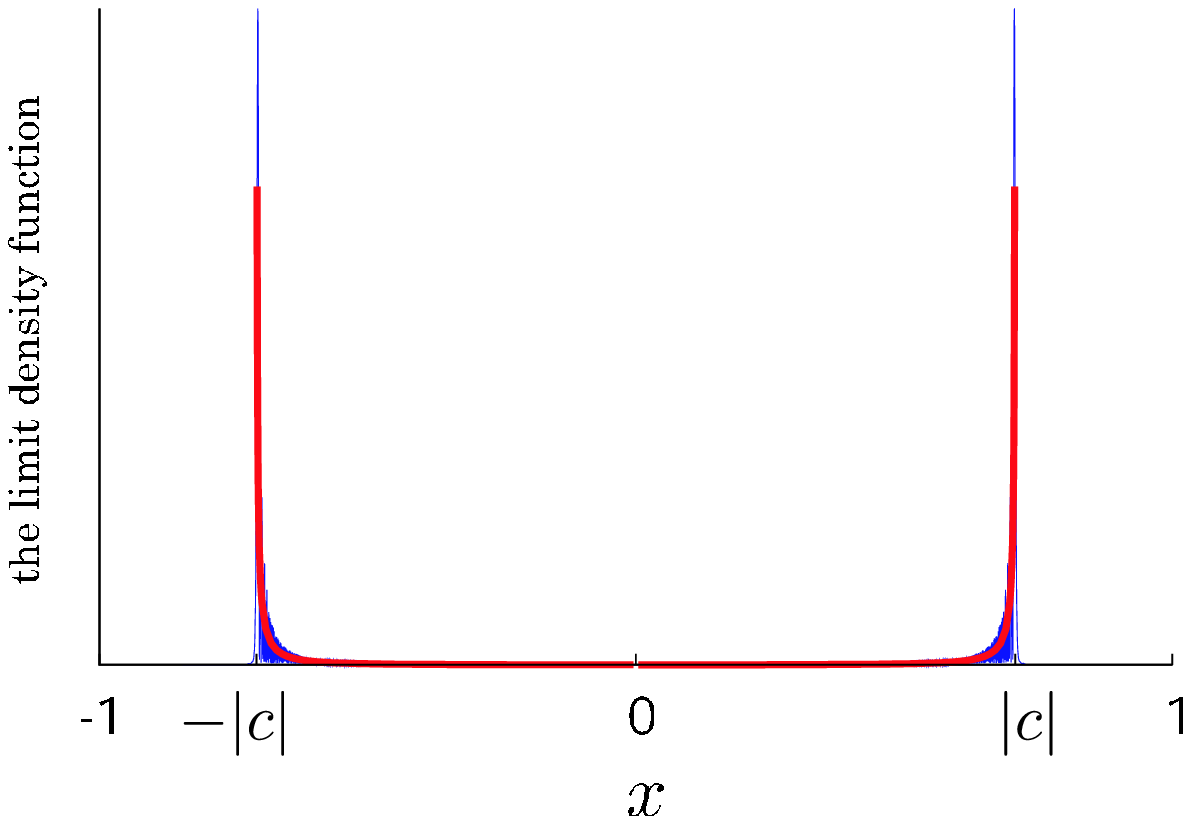}\\
    {(b) $\xi=1$}
   \end{center}
  \end{minipage}
 \end{center}
 \caption{(Case 3) Comparison between the limit density function (red line) and the probability distribution $\mathbb{P}(X_t/t=x)$ at time $t=5000$ (blue line) with $\alpha=1/\sqrt{2},\,\beta=i/\sqrt{2}$.}
  \label{fig:case3}
\end{figure}
\noindent Note that the limit density function with $\xi=0$ isn't continuous at $x=0$.
On the other hand, since $\lim_{x\to \pm|c|} f_1(x;0,\alpha,\beta)g_2(x;0)I_{(-|c|,|c|)}(x)=0$, it's continuous at $x=\pm |c|$.

\subsection{Case 4}
Let us suppose
\begin{align}
 w_1(k)=&\left\{\begin{array}{cl}
	  0&(-\pi \leq k < -\pi/2),\\[1mm]
		\sin k+1& (-\pi/2 < k \leq \pi/2),\\[1mm]
		 2& (\pi/2 \leq k <\pi),
		\end{array}\right.\\
 w_2(k)=&0.
\end{align}
These functions yield the following $(\alpha,\beta)$ delocalized initial state,
\begin{equation}
 \ket{\psi_0(x)}=\left\{\begin{array}{ll}
		  \frac{2}{\sqrt{7}}\ket{\phi}&(x=0), \\[1mm]
		  i\frac{2}{\sqrt{7}}\left(\frac{1}{4}+\frac{1}{\pi}\right)\frac{1}{x} \ket{\phi}&(x=\pm 1), \\[1mm]
		  i\frac{2}{\sqrt{7}\pi}\frac{1}{x} \ket{\phi}&(x=\pm 3,\pm 5,\ldots), \\[1mm]
		  i\frac{2}{\sqrt{7}\pi}\left(\frac{x}{x^2-1}-\frac{2}{x}\right) \ket{\phi}&(x=\pm 2,\pm 6,\ldots), \\[1mm]
		  -i\frac{2}{\sqrt{7}\pi}\frac{x}{x^2-1}\ket{\phi}&(x=\pm 4,\pm 8,\ldots).
			\end{array}\right.
\end{equation}
Then Eq.~(\ref{eq:limit_th}) has a representation
\begin{align}
 &\lim_{t\to\infty}\mathbb{E}\biggl[\biggl(\frac{X_t}{t}\biggr)^r\biggr]\nonumber\\
 =&\int_{-\infty}^{\infty}  x^r \bigl\{f_1(x;\xi,\alpha,\beta)g_3(x;\xi)+f_2(x;\xi,\alpha,\beta)g_4(x;\xi)\bigr\}I_{(-|c|,|c|)}(x)\,dx,\label{eq:limth_case4}
\end{align}
where
\begin{align}
 g_3(x;0)&=\frac{2}{7}\left\{3+\frac{c^2-x^2}{c^2(1-x^2)}\right\},\\
 g_3(x;1)&=\frac{2}{7}\left\{3+\frac{s^2x^2}{c^2(1-x^2)}\right\},\\
 g_4(x;0)&=\left\{\begin{array}{ll}
	    -\frac{8c}{7|c|}\left(1-\frac{\sqrt{c^2-x^2}}{|c|\sqrt{1-x^2}}\right)&(0\leq x< |c|),\\[1mm]
	    \frac{8c}{7|c|}\left(1-\frac{\sqrt{c^2-x^2}}{|c|\sqrt{1-x^2}}\right)&(-|c|<x<0),\\
		  \end{array}\right.\\
 g_4(x;1)&=\left\{\begin{array}{ll}
	    \frac{8}{7}\left(\frac{c}{|c|}-\frac{|s|x}{c\sqrt{1-x^2}}\right)&(0\leq x <|c|),\\[1mm]
	    -\frac{8}{7}\left(\frac{c}{|c|}+\frac{|s|x}{c\sqrt{1-x^2}}\right)&(-|c|< x< 0).	   
		  \end{array}\right.
\end{align}

\begin{figure}[h]
 \begin{center}
  \begin{minipage}{50mm}
   \begin{center}
    \includegraphics[scale=0.4]{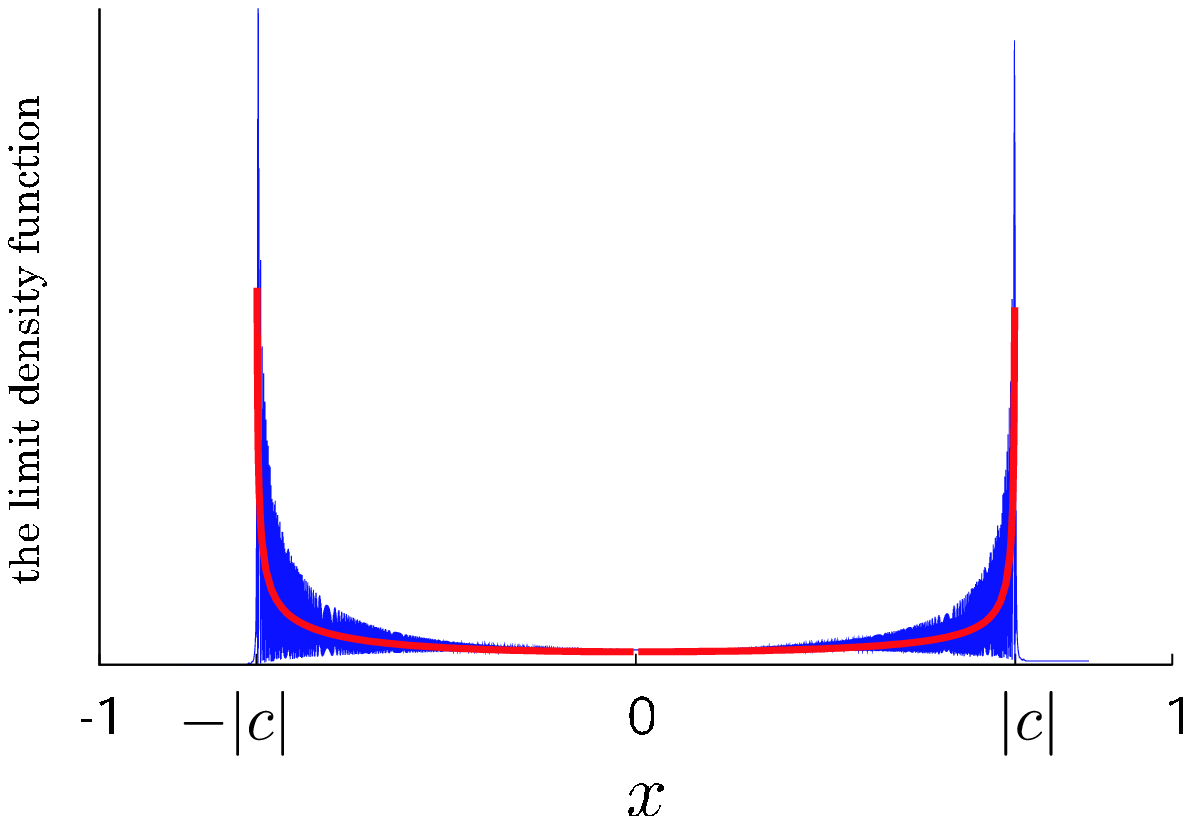}\\
    {(a) $\xi=0$}
   \end{center}
  \end{minipage}
  \begin{minipage}{50mm}
   \begin{center}
    \includegraphics[scale=0.4]{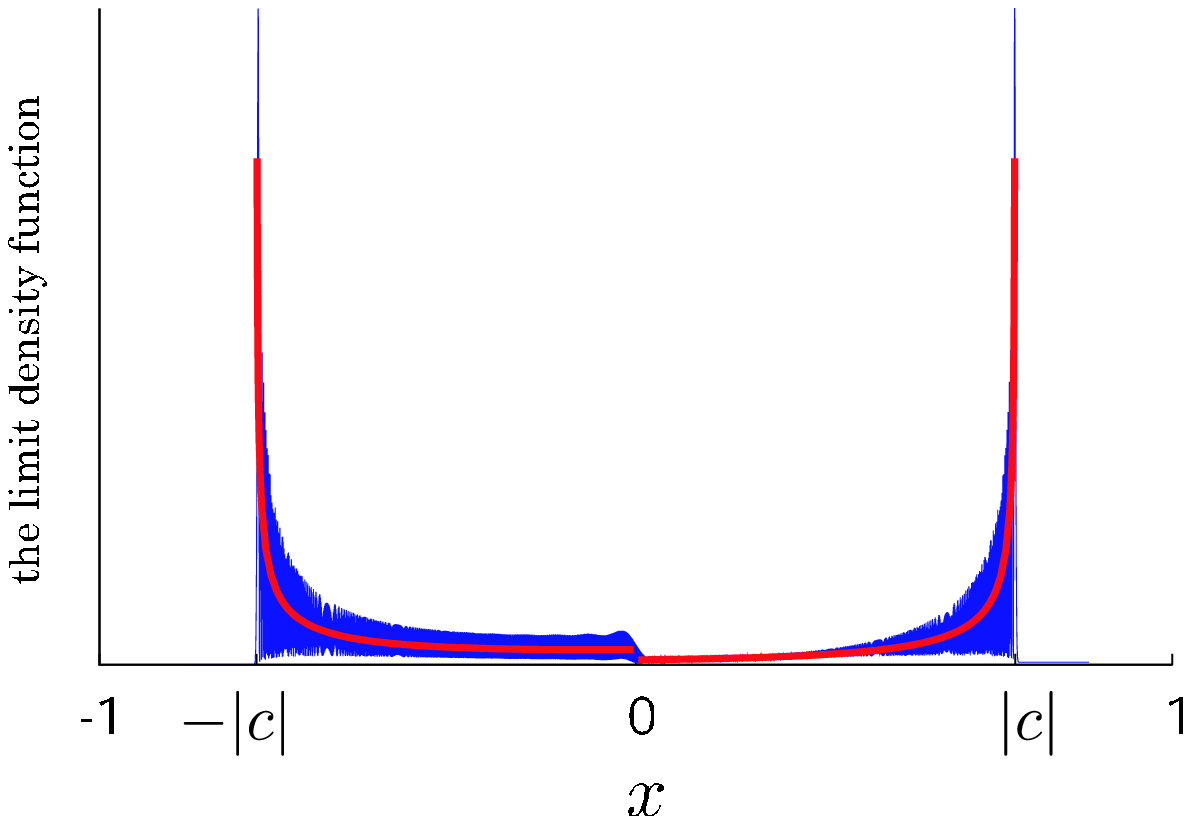}\\
    {(b) $\xi=1$}
   \end{center}
  \end{minipage}
 \end{center}
 \caption{(Case 4) Comparison between the limit density function (red line) and the probability distribution $\mathbb{P}(X_t/t=x)$ at time $t=5000$ (blue line) with $\alpha=1/\sqrt{2},\,\beta=i/\sqrt{2}$.}
  \label{fig:case4}
\end{figure}
\noindent Notice that the limit density function with $\xi=1$ isn't continuous at $x=0$.

\subsection{Case 5}
\begin{figure}[h]
 \begin{center}
  \begin{minipage}{50mm}
   \begin{center}
    \includegraphics[scale=0.4]{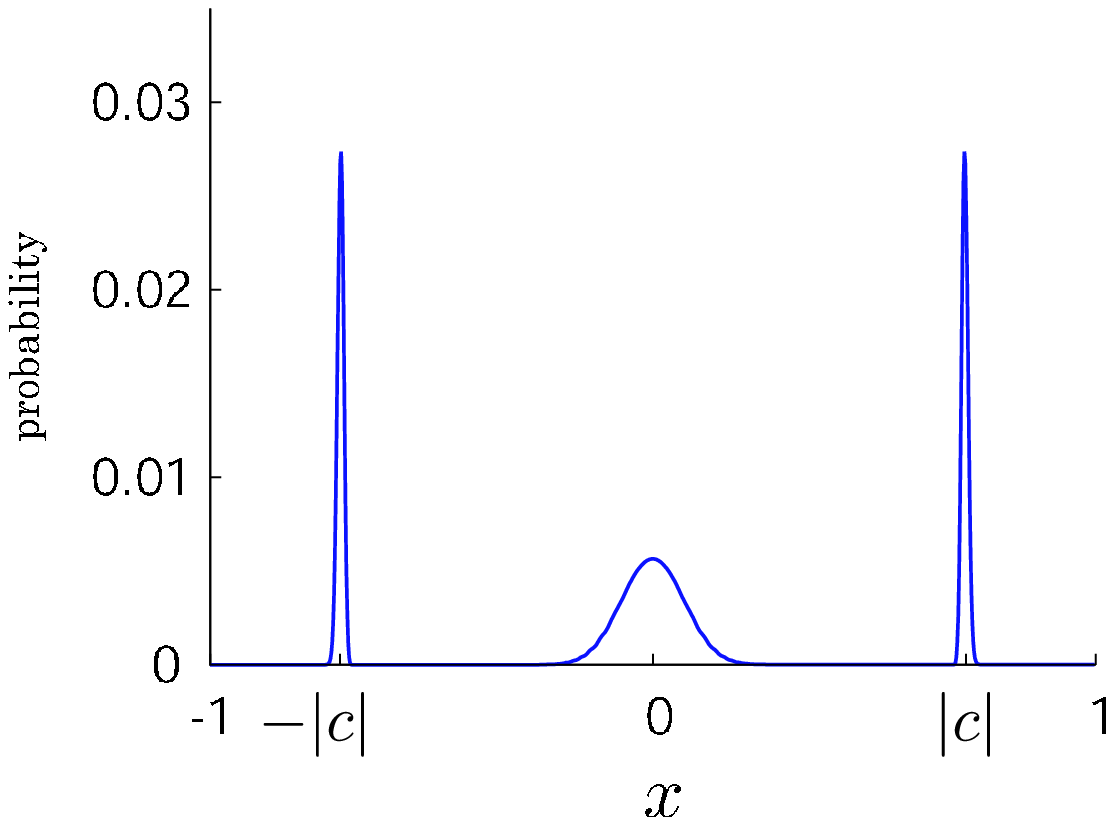}\\
    {(a) $\mathbb{P}(X_t/t=x)$}
   \end{center}
  \end{minipage}
  \begin{minipage}{50mm}
   \begin{center}
    \includegraphics[scale=0.4]{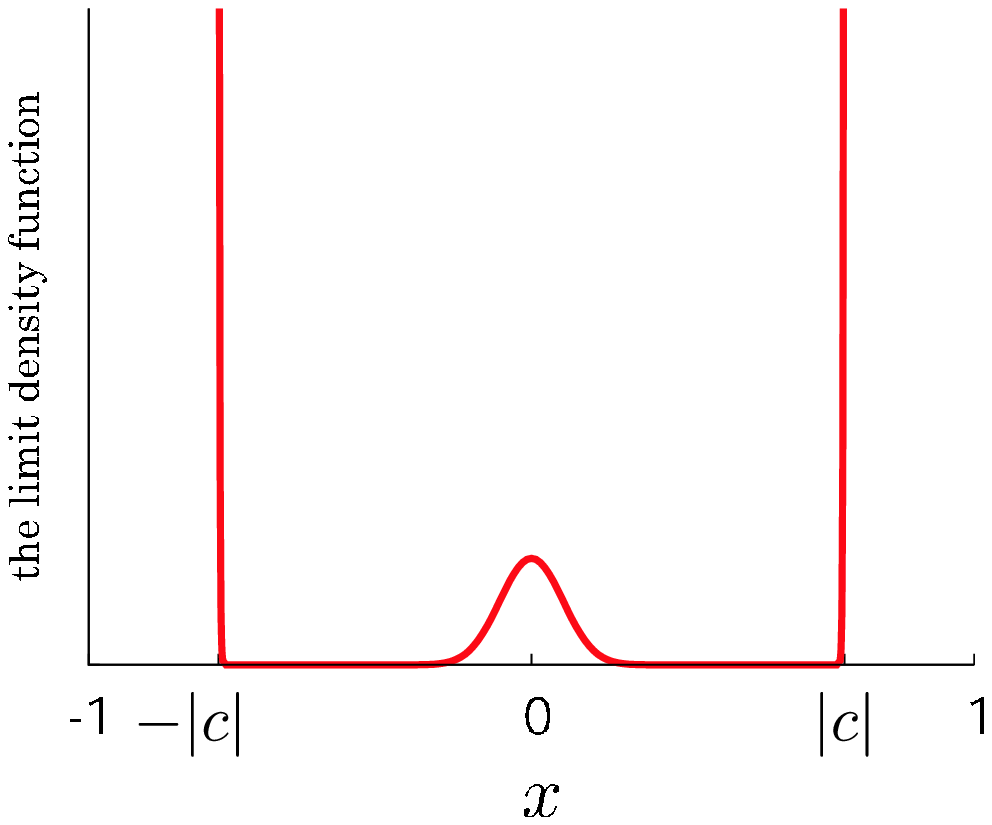}\\
    {(b) $f_1(x;\xi,\alpha,\beta)g_5(x)$}
   \end{center}
  \end{minipage}
 \end{center}
 \caption{(Case 5) Comparison between the probability distribution $\mathbb{P}(X_t/t=x)$ at time $t=1000$ (Fig.~(a)) and the limit density function (Fig.~(b)) with $n=50$ and $\alpha=1/\sqrt{2},\,\beta=i/\sqrt{2}$.}
  \label{fig:case5}
\end{figure}

As the final example, we focus on an initial state over a finite domain. 
For a non-negative integer $n$, we give
\begin{align}
 w_1(k)=&\cos^{2n+2}k+\sin^{2n+2}k,\\
 w_2(k)=&(-1)^n(\cos k\sin^{2n+1}k-\sin k\cos^{2n+1}k).
\end{align}
Then setting $\chi_{n,j}=(-1)^{n+1}\left\{(-1)^j(2j+1)-1\right\}\,(j=0,1,2,\ldots,n)$, we have the $(\alpha,\beta)$ delocalized initial state,
\begin{equation}
\ket{\psi_0(x)}=\left\{\begin{array}{ll}
		  \frac{1}{2^{2n}}\sqrt{\frac{\sqrt{\pi}\,\Gamma(2n+2)}{2\Gamma(2n+\frac{3}{2})}}{2n+1\choose n-j}\ket{\phi}&(\exists j,\, x=\chi_{n,j}),\\
			 0\ket{\phi}&(\forall j,\, x\neq \chi_{n,j}),
			\end{array}\right.
\end{equation}
where $\Gamma(z)=\int_0^\infty s^{z-1}e^{-s}\,ds$ is the Gamma function.
A limit distribution follows from Eq.~(\ref{eq:cor1}):
\begin{align}
 \lim_{t\to\infty}\mathbb{E}\biggl[\biggl(\frac{X_t}{t}\biggr)^r\biggr]=\int_{-\infty}^{\infty}& x^r f_1(x;\xi,\alpha,\beta)g_5(x)I_{(-|c|,|c|)}(x)\,dx,\label{eq:limth_case5}
\end{align}
where
\begin{align}
 g_5(x)&=\frac{\sqrt{\pi}\,\Gamma(2n+2)}{2\Gamma(2n+\frac{3}{2})}\frac{s^{4n+2}x^{4n+2}+(c^2-x^2)^{2n+1}}{c^{4n+2}(1-x^2)^{2n+1}}.
\end{align}
If $n=0$, then the limit distribution is equivalent to that of the QW starting from the origin~\cite{Konno2002,Konno2005}.
The effect of the parameters $\xi\in\left\{0,1\right\}$ is expressed in a linear term of $f_1(x;\xi,\alpha,\beta)$.
Figure~\ref{fig:case5} gives us a comparison between the probability distribution $\mathbb{P}(X_t/t=x)$ and the limit density function with $n=50$ and $\alpha=1/\sqrt{2}, \beta=i/\sqrt{2}$.

%\newpage

%%%%%%%%%%%%%%%%%%%   SUMMARY   %%%%%%%%%%%%%%%%%%%%%%%%%%%%%%%%%%%%
\section{Summary}
\label{sec:summary}
In this section, we will mention our results and a future problem.
Limit theorems of QWs have been mainly investigated for localized initial conditions.
Theoretical analysis for the QWs with a delocalized initial state is generally complex.
In this paper, to deepen our understanding of relationship between a 2-state QW on the line and its initial conditions, we produced the $(\alpha,\beta)$ delocalized initial state induced from the Fourier series expansion, and derived a long-time limit theorem for the walk.
That is, we found one of the QWs with a delocalized initial state whose approximate behavior after many steps can be forecasted.
By using our limit theorem, we have specifically computed limit density functions for some QWs with the $(\alpha,\beta)$ delocalized initial state.
As a result, Eqs.~(\ref{eq:limth_case2}), (\ref{eq:limth_case3}), (\ref{eq:limth_case4}) and (\ref{eq:limth_case5}) are different from the limit distribution for the 2-state QW starting with a localized initial state.
Compared to the QW with a localized initial condition, whose limit density function never has a discontinuity in the compact support, limit density functions of the QW with a delocalized initial condition can have a discontinuity in the compact support (See Figs.~\ref{fig:case2}-(a), \ref{fig:case3}-(a) and \ref{fig:case4}-(b)).  
The limit distributions, moreover, taught us the effects of the coin-sift operators $U_{\xi,\theta}=\cos\theta\ket{0}\bra{0}+(-1)^{\xi}\sin\theta\ket{0}\bra{1}+\sin\theta\ket{1}\bra{0}-(-1)^\xi\cos\theta\ket{1}\bra{1}\, (\xi\in\left\{0,1\right\})$ (See Figs.~\ref{fig:case2}, \ref{fig:case3} and \ref{fig:case4}).

As a development of our investigation, it is interesting to analyze long-time behavior of multi-state QWs with a delocalized initial state.
Since the multi-state QWs starting from the origin can localize in probability distributions after long time~\cite{InuiKonnoSegawa2005,InuiKonno2005,SegawaKonno2008}, it is a conceivable topic to clarify the difference between our 2-state QW and the multi-state QWs with a delocalized initial state.
Also, our result claims that we can create various probability distributions with a compact support from the 2-state QWs by controlling the functions $w_1(k),w_2(k)$.
If the QW after many steps is realized in a quantum system, we'll physically gain various probability distributions.
At the same time, that means we can obtain various random variables in quantum computing.
The author hopes that our results will contribute to the quantum computation theory.

%%%%%%%%%%%%%%%%   ACKNOWLEDGMENTS   %%%%%%%%%%%%%%%%%%%%%%%%%%%%%%%%%%%
The author acknowledges support from the Japan Society for the Promotion of Science and the Meiji University Global COE Program ``Formation and Development of Mathematical Sciences Based on Modeling and Analysis''.

%%%%%%%%%%%%%%%%%%%   BIBTEX   %%%%%%%%%%%%%%%%%%%%%%%%%%%%%%%%%%

%\bibliography{../bibtex/reference}

\begin{thebibliography}{10}

\bibitem{Konno2002}
N.~Konno,
\newblock {\em Quant. Inform. Comput.} {\bf 2}(Special Issue) (2002) 578.

\bibitem{Konno2005}
N.~Konno,
\newblock {\em J. Math. Soc. Jpn.} {\bf 57} (2005) 1179.

\bibitem{InuiKonnoSegawa2005}
N.~Inui {\em et al.},
\newblock {\em Phys. Rev. E} {\bf 72} (2005) 056112.

\bibitem{InuiKonno2005}
N.~Inui and N.~Konno,
\newblock {\em Phys. A} {\bf 353} (2005) 133.

\bibitem{SegawaKonno2008}
E.~Segawa and N.~Konno,
\newblock {\em Int. J. Quant. Inform.} {\bf 6} (2008) 1231.

\bibitem{KonnoMachida2010}
N.~Konno and T.~Machida,
\newblock {\em Quant. Inform. Comput.} {\bf 10} (2010) 1004.

\bibitem{WatabeKobayashiKatoriKonno2008}
K.~Watabe {\em et al.},
\newblock {\em Phys. Rev. A} {\bf 77} (2008) 062331.

\bibitem{DiMcMachidaBusch2011}
C.~Di~Franco {\em et al.},
\newblock {\em Phys. Rev. A} {\bf 84} (2011) 042337.

\bibitem{ChisakiHamadaKonnoSegawa2009}
K.~Chisaki {\em et al.},
\newblock {\em Int. Inform. Sci.} {\bf 15} (2009) 423.

\bibitem{ChisakiKonnoSegawa2010}
K.~Chisaki {\em et al.},
\newblock {\em Quant. Inform. Comput.} {\bf 12} (2012) 314.

\bibitem{KonnoObataSegawa2013}
N.~Konno {\em et al.},
\newblock {\em Comm. Math. Phys.} {\bf 322} (2013) 667.

\bibitem{LiuPetulante2012}
C.~Liu and N.~Petulante,
\newblock {\em Int. J. Quant. Inform.} {\bf Online Ready} (2013) 1350054.


\bibitem{AbalDonangeloRomanelliSiri2006}
G.~Abal {\em et al.},
\newblock {\em Phys. A} {\bf 371} (2006) 1.

\bibitem{AbalSiriRomanelliDonangelo2006}
G.~Abal {\em et al.},
\newblock {\em Phys. Rev. A} {\bf 73} (2006) 042302.

\bibitem{ValcarcelRoldanRomanelli2010}
G.J.~Valc{\'a}rcel {\em et al.},
\newblock {\em New Journal of Physics} {\bf 12} (2010) 123022.

\bibitem{ChandrashekarBusch2012}
C.~M.~Chandrashekar and Th.~Busch,
\newblock {\em Quant. Inform. Process.} {\bf 11} (2012) 1287.

\bibitem{BouwmeesterMarzoliKarmanSchleichWoerdman1999}
D.~Bouwmeester {\em et al.},
\newblock {\em Phys. Rev. A} {\bf 61} (1999) 013410.

\bibitem{BroomeFedrizziLanyonKassalAspuru-GuzikWhite2010}
M.A.~Broome {\em et al.},
\newblock {\em Phys. Rev. Lett.} {\bf 104} (2010) 153602.

\bibitem{DoStohlerBalasubramanianElliottEashFischbachFischbachMillsZwickl2005}
B.~Do {\em et al.},
\newblock {\em JOSA B} {\bf 22} (2005) 499.

\bibitem{KarskiForsterChoiSteffenAltMeschedeWidera2009}
M.~Karski {\em et al.},
\newblock {\em Science} {\bf 325} (2009) 174.

\bibitem{PeretsLahiniPozziSorelMorandottiSilberberg2008}
H.B.~Perets {\em et al},
\newblock {\em Phys. Rev. Lett.} {\bf 100} (2008) 170506.

\bibitem{SchmitzMatjeschkSchneiderGlueckertEnderleinHuberSchaetz2009}
H.~Schmitz {\em et al.},
\newblock {\em Phys. Rev. Lett.} {\bf 103} (2009) 90504.

\bibitem{SchreiberCassemiroPotovcekGabrisMosleyAnderssonJexSilberhorn2010}
A.~Schreiber {\em et al.},
\newblock {\em Phys. Rev. Lett.} {\bf 104} (2010) 50502.

\bibitem{ZahringerKirchmairGerritsmaSolanoBlattRoos2010}
F.~Z{\"a}hringer {\em et al.},
\newblock {\em Phys. Rev. Lett.} {\bf 104} (2010) 100503.

\bibitem{MeineckePouliosPolitiMatthewsPeruzzoIsmailWorhoffOBrienThompson2013}
Jasmin D.~A.~Meinecke {\em et al.},
\newblock {\em Phys. Rev. A} {\bf 88} (2013) 012308.

\bibitem{SansoniSciarrinoValloneMataloniCrespiRamponiOsellame2012}
L.~Sansoni {\em et al.},
\newblock {\em Phys. Rev. Lett.} {\bf 108} (2012) 10502.

\bibitem{GrimmettJansonScudo2004}
G.~Grimmett {\em et al.},
\newblock {\em Phys. Rev. E} {\bf 69} (2004) 026119.

\bibitem{Machida2011}
T.~Machida,
\newblock {\em Int. J. Quant. Inform.} {\bf 9} (2011) 863.

\bibitem{ChisakiKonnoSegawaShikano2011}
K.~Chisaki {\em et al.},
\newblock {\em Quant. Inform. Comput.} {\bf 11} (2011) 741.

\bibitem{JorsboeMejlbro1982}
O.~J{\o}rsboe and L.~Mejlbro,
\newblock {\em The Carleson-Hunt theorem on fourier series}, Vol. 911 (Springer-Verlag, 1982).

\bibitem{Reyna2002}
J. A.~De Reyna,
\newblock {\em Pointwise convergence of Fourier series}, Vol. 1785 (Springer-Verlag, 2002).

\end{thebibliography}

\end{document}